\documentclass
[aps,prd,twocolumn,superscriptaddress,nofootinbib]{revtex4-1}%
\usepackage{amsmath}
\usepackage{amsfonts}
\usepackage{amssymb}
\usepackage{graphicx}
\usepackage{latexsym}
\usepackage{color,epsfig}
\usepackage{ifpdf}
\usepackage{bm}
\usepackage{wasysym}
\usepackage[english]{babel}
\usepackage{braket}
\usepackage{slashed}
\usepackage{longtable,array,cancel}

\definecolor{nicered}{rgb}{0.7,0.1,0.1}
\definecolor{nicegreen}{rgb}{0.1,0.5,0.1}

\newcommand{\gs}[1]{\cancel{#1}}

\newcommand{\op}{{\cal Q}}
\newcommand{\nn}{\nonumber}
\newcommand{\ii}{\mathrm{i}}
\begin{document}

\title{Interplay of $t \to b W$ decay and $B_{d,s} - \bar{B}_{d,s}$ mixing in Minimal Flavor Violating Models
}

\author{Jure Drobnak}
\email[Electronic address:]{jure.drobnak@ijs.si}
\affiliation{J. Stefan Institute, Jamova 39, P. O. Box 3000, 1001  Ljubljana, Slovenia} 

\author{Svjetlana Fajfer} 
\email[Electronic address:]{svjetlana.fajfer@ijs.si} 
\affiliation{J. Stefan Institute, Jamova 39, P. O. Box 3000, 1001
  Ljubljana, Slovenia}
\affiliation{Department of Physics,
  University of Ljubljana, Jadranska 19, 1000 Ljubljana, Slovenia}

\author{Jernej F. Kamenik}
\email[Electronic address:]{jernej.kamenik@ijs.si} 
\affiliation{J. Stefan Institute, Jamova 39, P. O. Box 3000, 1001  Ljubljana, Slovenia}
\affiliation{Department of Physics,
  University of Ljubljana, Jadranska 19, 1000 Ljubljana, Slovenia}

\date{\today}

\begin{abstract}
Precise measurements of the top quark decay properties at hadron colliders offer interesting new possibilities of testing the standard model. At the same time, recent intriguing experimental results concerning CP violation in the $B_d$ and $B_s$ systems have stimulated many studies of physics beyond the standard model. 
We investigate anomalous $t W d_j$ interactions as a possible source of new effects in $B_{d,s}- \bar B_{d,s}$ oscillations within a model independent approach based on the assumptions of Minimal Flavor Violation. After matching
our effective operators onto the low-energy effective Lagrangian describing $B_{d,s}$ meson mixing and evolving it down to the $B$-mass scale, we extract the preferred ranges of the anomalous $t W d_j$ interactions at the weak scale. These values are then compared to previously considered constraints coming from the rare radiative $B \to X_s \gamma$ decay. Finally, we reconsider the associated effects in the $t \to b W$ decays and find that the $W$ helicity fractions ${\cal F}_{L,+}$ can deviate by as much as $15\%$, $30\%$ from their standard model values, respectively.  The deviations in ${\cal F}_{L}$ in particular, can reach the level of expected precision measurements at the LHC.
\end{abstract}

\maketitle

\section{Introduction}
The extensive production of top quarks at the LHC and Tevatron colliders offers the possibility to study $tWb$ interactions with high accuracy. In particular, in the recent years the CDF and D0 collaborations have been measuring the helicity fractions of the $W$ boson from top quark decays with increasing precision~\cite{tbWexp}. 

Motivated by these results, we have recently considered contributions of effective operators of dimension six or less~\cite{AguilarSaavedra:2008zc} contributing at tree level to the decay of an unpolarized top quark to a bottom quark and a $W$ gauge boson at next-to-leading order (NLO) in QCD \cite{Drobnak:2010ej}. Effects of such operators are suppressed within the standard model (SM) but might be enhanced by new physics (NP) contributions. We found that NLO QCD effects can be significant for the transverse-plus helicity fraction (${\cal F}_+$) of the $W$ by lifting the helicity suppression of the operators containing left-handed bottom quark fields.
While it turns out that indirect constraints from radiative $B\to X_s\gamma$ decay~\cite{Grzadkowski:2008mf} already severely constrain possible NP effects in this observable, we found that the CDF measurement of the longitudinal helicity fraction (${\cal F}_L$) of the $W$ boson already provides the most stringent upper bound on the contributions of one of the NP operators~\cite{Drobnak:2010ej}.

In addition to the effects in top quark decays mentioned above, anomalous $tWd_j$ interactions, with $j$ labeling quark flavor, might manifest themselves in rare flavor changing neutral current processes of $B$ mesons where loops involving top quarks play a crucial role.  As already mentioned, such effects have been analyzed in the radiative $B\to X_s \gamma$ decay~\cite{Grzadkowski:2008mf}
 resulting in restrictive bounds on effective operator contributions describing possible deviations in the $t\to b W$ decay properties from the SM. 
  
Recently, possible NP effects in the $B_{d,s}-\bar B_{d,s}$, mixing amplitudes have received considerable attention (c.f.~\cite{Lenz:2010gu} and references within). In particular within the SM, the $B^0-\bar B^0$ mass difference and the time-dependent CP asymmetry in $B\to J/\psi K_s$ are strongly correlated with the branching ratio $\mathrm{Br}(B^+\to \tau^+ \nu)$~\cite{Deschamps:2008de}. The most recent global analyses point to a disagreement of this correlation with direct measurements at the level of 2.9 standard deviations~\cite{Lenz:2010gu}. Similarly in the $B_s$ sector the recently measured CP-asymmetries by the Tevatron experiments, namely in $B_s \to J/\psi\phi$~\cite{betas} and in di-muonic inclusive decays~\cite{Abazov:2010hj} when combined, deviate from the SM prediction for the CP violating phase in $B_s-\bar B_s$ mixing by 3.3 standard deviations~\cite{Lenz:2010gu}. 

In the present paper we consider anomalous $tWd_j$ interactions as a possible solution of these anomalies via their contribution to $B_{d,s}-\bar B_{d,s}$ mixing amplitudes. Similar analyses have been attempted recently~\cite{Lee:2008xs,Lee:2010hv} using a subset of all possible effective $tWd_j$ operators. In particular, the authors of~\cite{Lee:2008xs} have constrained contributions of two possible $tWb$ operators using mostly $B_d$ oscillation parameter measurements,  while in Ref.~\cite{Lee:2010hv} they  considered two $tWs$ operators and concluded that the large non standard CP violating phase in $B_s-\bar B_s$ mixing can be accommodated without violating $B\to X_s\gamma$ constraints provided that the two NP operator contributions conspire in a way to mostly cancel in the later process while at the same time significantly affect the former. 



An important issue to consider when interpreting indirect bounds on effective operators concerns the choice of flavor basis for the operators. Since the SM electro-weak symmetry breaking (EWSB) induces misalignment between the up and down quark mass eigenbasis and the weak interaction eigenbasis via the CKM mechanism, isolating NP effects in $tWd_j$ interactions to a particular single (e.g. $t\to s$) flavor transition in the physical (mass) basis in general requires a large degree of fine-tuning in the flavor structure of the effective operators in the UV. This is the case in the existing analysis of anomalous $tWs$ effects in $B_s$ oscillations~\cite{Lee:2010hv}. One possible solution is to require the operators to be diagonal (aligned) in the weak interaction basis as done for example in the $B\to X_s\gamma$ analysis~\cite{Grzadkowski:2008mf}. Then, EWSB generically induces several $u_iWd_j$ flavor transitions, whose relative strengths are governed by the CKM matrix elements, resulting effectively in minimal flavor violating (MFV) scenarios~\cite{MFV}. This is also the approach we employ in this work, paying however special attention to the possible controlled breaking of such flavor universality by large bottom yukawa effects and their implications for new CP violating effects in $B_{d,s}-\bar B_{d,s}$ mixing as recently discussed in Ref.~\cite{GMFV}. Finally, we focus our discussion on the operators which can be probed directly in the $t\to b W$ decay measurements, since this provides an exciting possibility to directly test the suggested origin of new CP violating sources in the $B_{d,s}$ systems at high energy colliders.

This paper is structured as follows. First we introduce the model independent approach employed and define our effective Lagrangian including the (MFV) flavor structure of the NP operators mediating $t\to b W$ transitions above the EWSB scale. In the main part of the paper we describe the matching computation for passing from such effective theory to another low-energy effective theory relevant for $B_{d,s}$ meson mixing where the top quark and the electroweak gauge bosons have already been integrated out. Our results are then combined with the recent global CKM and $B_{d,s}$ mixing fits to extract preferred ranges for the anomalous $tWd_j$ interactions and give corresponding predictions for the $W$ helicity fractions in the $t\to b W$ decays. We conclude our work in the last section.

\section{Framework}
We work in the framework of an effective theory, described by the Lagrangian
\begin{eqnarray}
{\cal L}={\cal L}_{\mathrm{SM}}+\frac{1}{\Lambda^2}\sum_i C_i \mathcal Q_i +\mathrm{h.c.}+ {\cal O}(1/\Lambda^3)\,,
\label{eq:lagr}
\end{eqnarray}
where ${\cal L}_{\mathrm{SM}}$ is the SM part, $\Lambda$ is the scale of NP and $\mathcal Q_i$ are dimension-six operators, invariant under SM gauge transformations and consisting of SM fields. {In doing this we assume that at the scale $m_t$ the SM fields with up to two Higgs doublets are the only propagating degrees of freedom, that the electroweak symmetry is only broken by the vacuum expectation values of these two scalars and that operators up to dimension six give the most relevant contributions to the observables we consider.} Such an approach is appropriate to summarize weak scale effects of NP at $\Lambda \gg m_t$, where the new heavy degrees of freedom have been integrated out.

Our operator basis consists of all dimension-six operators that generate charged current quark interactions with the $W$. Since we restrict our discussion to MFV scenarios, Lagrangian (\ref{eq:lagr}) has to be formally invariant under the SM flavor group $\mathcal G^{\rm SM} = U(3)_{Q} \times U(3)_u \times U(3)_d$ where $Q, u, d$ stand for quark doublets and up and down type quark singlets respectively. MFV requires that the only $\mathcal G^{\rm SM}$ symmetry breaking spurionic fields in the theory are the up and down quark Yukawa matrices $Y_{u,d}$, formally transforming as $(3,\bar 3,1)$ and $(3,1,\bar 3)$ respectively. 

We identify four relevant quark bilinears with distinct transformation properties under $\mathcal G^{\rm SM}$: $\bar u d$, $\bar Q Q$, $\bar Q u$ and $\bar Q d$ transforming as $(1,\bar 3, 3)$, $({1\oplus8},1,1)$, $(\bar 3,3,1)$ and $(\bar 3, 1, 3)$ respectively. Using these, we can construct the most general $\mathcal G^{\rm SM}$ invariant quark bilinear flavor structures as
\begin{equation}
\bar u Y_u^\dagger \mathcal A_{ud} Y_d d\,, ~~~ \bar Q \mathcal A_{QQ} Q\,, ~~~ \bar Q \mathcal A_{Qu} Y_u u\,,~~~ \bar Q \mathcal A_{Qd} Y_d d\,,
\label{eq:flav}
\end{equation}
where $\mathcal A_{xy}$ are arbitrary polynomials of $Y_{u} Y_{u}^\dagger$ and/or $Y_{d}Y_d^\dagger$,  transforming as $({1\oplus8},1,1)$.

In order to identify the relevant flavor structures in terms of physical parameters, we can without the loss of generality consider $Y_{u,d}$ condensate values in a basis in which $\langle Y_d \rangle$ is diagonal: $\langle Y_d \rangle =\mathrm{diag}(m_d,m_s,m_b)/v_d$ and $\langle Y_u \rangle =V_{}^\dagger \mathrm{diag}(m_u,m_c,m_t)/v_u$, where we have introduced separate up- and down-type Higgs condensates $v_{u,d}$, while $V_{}$ is the SM CKM matrix. We also write $Q,u,d$ in this basis in terms of  quark mass eigenstates $u_{Li}, d_{Li}, u_{Ri}, d_{Ri}$ as $Q_i=(V^*_{ki} u_{Lk},d_{Li})$, $u_i=u_{iR}$ and $d_i=d_{iR}$, where $L,R$ subscripts denote chirality projectors $\psi_{R,L} = (1\pm \gamma_5) \psi/2$. 

We consider first the simplest case of linear MFV where within $\langle \mathcal A_{xy} \rangle$ higher powers of $\langle Y_d Y_d^\dagger\rangle \simeq \mathrm{diag}(0,0,m^2_b/v^2_d) $ can be neglected. Neglecting also contributions suppressed by first and second generation quark masses, the only relevant flavor contributions of the arbitrary $\mathcal A_{xy}$ structures in Eq.~(\ref{eq:flav}) are 
\begin{align}
\bar t_R V_{tb}  b_R\,, && \bar Q_i Q_i\,, && \bar Q_i V^*_{ti} V_{tj} Q_j\,, \nonumber\\
\bar Q_i V^*_{ti} t_R\,, && \bar Q_3  b_R\,, && \bar Q_i V^*_{ti} V_{tb} b_R\,,
\end{align}
where summation over repeated $(i,j)$ flavor indices is understood.
{We note in passing that within NP models where higher powers of Yukawa insertions are suppressed by small perturbative parameters, the structures $\bar Q_i V^*_{ti} V_{tj} Q_j$ and $\bar Q_i V^*_{ti} V_{tb} b_R$ are subleading compared to $ \bar Q_i Q_i$ and $\bar Q_3  b_R$ respectively.}
Notice that since $\bar Q_i Q_i$ is completely flavor universal, when coupled to the $W$ it would modify the effective Fermi constant as extracted from charged quark currents compared to the muon lifetime. Existing tight constraints on such deviations~\cite{Antonelli:2009ws} do not allow for significant effects in $B_{d,s}$ or top quark phenomenology and we do not consider this structure in our analysis. On the other hand, $\bar Q_i V^*_{ti} V_{tj} Q_j$ potentially leads to large tree level flavor changing neutral currents (FCNCs) in the down quark sector if coupled to the $Z$. This restricts the $SU(2)_L$ structure of such an operator in order to contribute significantly to charged current interactions~\cite{Grzadkowski:2008mf}.  Similarly, $\bar Q_i V^*_{ti} V_{tb} b_R$ if coupled to the photon or $Z$  would generate large tree level FCNC $\Delta B=1$ transitions, which are already tightly constrained by $B\to X_s\gamma$ and $B\to X_s \ell^+ \ell^-$~\cite{Hurth:2008jc}. Since all the charged current mediating $SU(2)_L$ invariant operators of dimension six or less containing such a flavor structure do necessarily involve either the $Z$ or the photon, we drop this structure from our subsequent analysis.

Taking these considerations into account, we finally obtain the following relevant set of dimension six $SU(2)_L$ invariant effective operators mediating charged quark currents in linear MFV NP scenarios
\begin{eqnarray}
 \mathcal Q_{RR}&=& V_{tb} [\bar{t}_R\gamma^{\mu}b_R] \big(\phi_u^\dagger\ii D_{\mu}\phi_d\big) \,, \nn\\
\nn \mathcal Q_{LL}&=&[\bar Q^{\prime}_3\tau^a\gamma^{\mu}Q'_3] \big(\phi_d^\dagger\tau^a\ii D_{\mu}\phi_d\big) \hspace{-0.1cm}-\hspace{-0.1cm}[\bar Q'_3\gamma^{\mu}Q'_3]\big(\phi_d^\dagger\ii D_{\mu}\phi_d\big),\\
\nn \mathcal Q_{LRt} &=& [\bar Q'_3 \sigma^{\mu\nu}\tau^a t_R]{\phi_u}W_{\mu\nu}^a \,,\\
 \mathcal Q_{LRb} &=& [\bar Q_3 \sigma^{\mu\nu}\tau^a b_R]\phi_d W_{\mu\nu}^a \,,\label{operators}
\label{eq:ops1}
\end{eqnarray}
where we have introduced $\bar Q'_3 =\bar Q_i V^*_{ti}= (\bar{t}_L,V_{ti}^*\bar{d}_{iL})$, $\sigma^{\mu\nu}=\ii [\gamma^{\mu},\gamma^{\nu}]/2$, $W^a_{\mu\nu}=\partial_{\mu}W_{\nu}^a-\partial_{\nu}W_{\mu}^a - g\epsilon_{abc}W_{\mu}^b W_{\nu}^c$, and $\phi_{u,d}$ are the up- and down-type Higgs fields (in the SM $\phi_u =\ii \tau^2 \phi_d^*$).  The final set of operators coincides with those considered in the $B\to X_s \gamma$ analysis~\cite{Grzadkowski:2008mf}. 
{Within the formal MFV expansion, the operators $\mathcal Q_{RR}$ and $\mathcal Q_{LRb}$ involving the right-handed $b$ quarks always appear with a pre-factor of the third eigenvalue of $Y_d$ ($m_b/v_d$). Although not written out explicitly in (\ref{eq:ops1}) these operators are therefore expected to be parametrically suppressed compared to $\mathcal Q_{LL}$ and $\mathcal Q_{LRt}$.}
Notice that starting with the most general MFV construction we are led to a set of operators, where largest deviations in charged quark currents are expected to involve the third generation (a notable exception being the flavor universal $\bar Q_i Q_i$ structure present already in the SM). 

Following~\cite{GMFV}, the generalization of the above discussion to MFV scenarios where large bottom Yukawa effects can be important is straight forward. Higher powers of $\langle Y_d Y_d^\dagger\rangle$ within $\mathcal A_{xy}$ effectively project to the third generation in the down sector yielding the following additional flavor structures
\begin{align}
\bar Q_3 Q_3\,, && \bar Q_3 V^*_{tb} V_{tj} Q_j\,, && \bar Q_3 V^*_{tb} t_R\,.
\end{align}
{Again note that in NP models where higher powers of Yukawa insertions are suppressed by small perturbative parameters, these structures are subleading compared to $\bar Q_i Q_i$, $\bar Q_i V^*_{ti} V_{tj} Q_j$ and $\bar Q_i V^*_{ti} t_R$ respectively.}
Now $ \bar Q_3 V^*_{tb} V_{tj} Q_j$ contains both a flavor universal charged current contribution as well as a FCNC structure which we both remove using a suitable $SU(2)_L$ assignment in the effective operator. Similarly, $\bar Q_3 Q_3$ structure, coupled to the $Z$ would contribute to the $Z\to b\bar b$ decay branching ratio, which in excellent agreement with the SM prediction at the $3\permil$ level~\cite{:2005ema}. Thus we also remove such contributions via suitable $SU(2)_L$ assignments leading altogether to three distinctly new operators
\begin{eqnarray}
\nn \mathcal Q'_{LL}&=&[\bar Q_3\tau^a\gamma^{\mu}Q_3] \big(\phi_d^\dagger\tau^a\ii D_{\mu}\phi_d\big) \hspace{-0.1cm}-\hspace{-0.1cm}[\bar Q_3\gamma^{\mu}Q_3]\big(\phi_d^\dagger\ii D_{\mu}\phi_d\big),\\
\nn \mathcal Q^{\prime\prime}_{LL}&=&  V_{tb}^*  \left\{[\bar Q'_3\tau^a\gamma^{\mu}Q_3] \big(\phi_d^\dagger\tau^a\ii D_{\mu}\phi_d\big) \right. \\
\nn&&\left.- [\bar Q'_3\gamma^{\mu}Q_3]\big(\phi_d^\dagger\ii D_{\mu}\phi_d\big)\right\} ,\\
 \mathcal Q'_{LRt} &=& V_{tb}^* \, [\bar Q_3 \sigma^{\mu\nu}\tau^a t_R]{\phi_u}W_{\mu\nu}^a \,.\label{operators}
\label{eq:ops2}
\end{eqnarray}
The most important effect of large bottom Yukawa contributions (i.e. appearance of operators $\mathcal Q''_L$ and $\mathcal Q'_{LRt}$) is that modifications of $tWb$ interactions with left-handed $t$ and/or $b$ quarks ({\it{written in the physical (mass) quark basis}}) can effectively be decoupled from those involving the first two generations. 

This completes our operator construction and after EWSB we obtain Feynman rules with anomalous $u_iWd_j$ couplings relevant in MFV scenarios. These are presented in the Appendix~\ref{sec:App1}, where we have chosen the SM normalization of the two Higgs condensates. Note also that, contrary to \cite{Grzadkowski:2008mf}, our operators are not Hermitian and we allow the Wilson coefficients to be complex.

\section{Matching}
In order to study the effects of the anomalous $u_i W d_j$ interactions on the matrix elements relevant in $B_{d,s}-\bar B_{d,s}$ mixing, we normalize them to the SM values by writing~\cite{Lenz:2010gu} 
\begin{equation}
M_{12}^{(d,s)}=\langle \bar{B}_{d,s}^0|{\cal H}_{\mathrm{eff}}|B_{d,s}^0\rangle/2m_{B_{d,s}} = M_{12}^{\mathrm{SM}(d,s)}\Delta_{d,s}\,,\label{mat}
\end{equation}
where $\Delta_{d,s}\neq 1$ signals NP contributions. We match our effective theory (\ref{eq:lagr}) at leading order (LO) in QCD to a new low-energy effective theory relevant for $\Delta B=2$ transitions and governed by the Lagrangian 
\begin{eqnarray}
{\cal L}_{\mathrm{eff}} =-\frac{G_F^2 m_W^2}{4 \pi^2}\big(V_{tb}V^*_{td,s}\big)^2 \sum_{i=1}^5 C_i(\mu) {\cal Q}_i^{d,s}\,,
\label{eq:lagrBB}
\end{eqnarray}
{where the operator basis is given in \cite{Becirevic:2001jj}. As will become evident below, the only relevant operators for our analysis are
${\cal Q}_1^d = [\bar{d}_L^{\alpha}\gamma^{\mu}b^{\alpha}_L][\bar{d}_L^{\beta}\gamma_{\mu}b_{L}^{\beta}]$ and 
${\cal Q}_1^s = [\bar{s}_L^{\alpha}\gamma^{\mu}b^{\alpha}_L][\bar{s}_L^{\beta}\gamma_{\mu}b_{L}^{\beta}]$, where $\alpha$ and $\beta$ are color indices.}

\begin{figure}[h]
\begin{center}
\includegraphics[scale= 0.6]{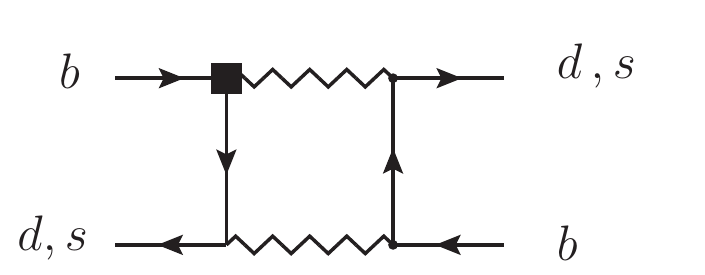}
\caption{Box diagram for $B_{d,s}-\bar{B}_{d,s}$ mixing. Square labels an anomalous coupling originating from one of the operators of Eq.~(\ref{eq:ops1},\ref{eq:ops2}). The zigzag lines represent $W$ gauge bosons or would-be Goldstone scalars $\phi$. Quarks running in the loop are up-type quarks.}
\label{slika1}
\end{center}
\end{figure}

In the matching procedure the $W$ boson and the top quark are integrated out by computing the box diagrams such as the one depicted in Fig.~\ref{slika1}. Diagrams where the anomalous couplings appear in the bottom-right corner instead the top-left and the crossed diagrams with internal quark and boson lines exchanged are completely symmetric and need not be computed separately. 

The calculation is done in the limit of massless external states and in a general $R_{\xi}$ gauge for the weak interactions. This allows us to verify gauge invariance of our final results. The drawback is the appearance of would-be Goldstone contributions including their new interactions generated by $\mathcal Q^{({\prime},{\prime}{\prime})}_{LL}$ and $\mathcal Q_{RR}$ operators.

The effective operators in Eq.~(\ref{eq:ops1}, \ref{eq:ops2}) in general generate several interaction vertices relevant for $B_{d,s}-\bar B_{d,s}$ oscillations: while $\mathcal Q_{RR}$, $\mathcal Q'_{LRt}$ and $\mathcal Q''_{LL}$ only modify the $tWb$ vertex,  $\mathcal Q_{LL}$ and $\mathcal Q_{LRt}$ also modify $tWs$ and $tWd$. Finally, $\mathcal Q_{LRb}$ and $\mathcal Q'_{LL}$ modify $tWb$ , but also $uWb$ and $cWb$.  Consequently, $\mathcal Q_{LL}$ and $\mathcal Q_{LRt}$ also contribute to $K^0-\bar K^0$ mixing at one-loop. In fact, their contributions to neutral kaon and $B$ meson oscillations turn out to be universal and purely real (see discussion below Eq.~(\ref{eq:kappas})). On the other hand, $\mathcal Q_{LRb}$ and $\mathcal Q'_{LL}$ could interfere with $V_{cb}$ and $V_{ub}$ extraction from semileptonic $B$ decays. Since these quantities are crucial for the reconstruction of the CKM matrix in MFV models, a consistent analysis of these operators would  require a modified CKM unitarity fit, which is beyond the scope of this paper and we leave it for a future study. 

\begin{figure}[h]
\begin{center}
\includegraphics[scale= 0.6]{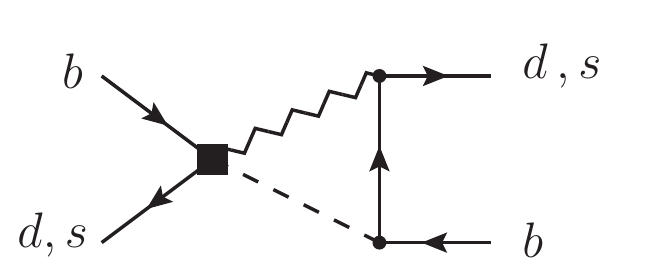}
\caption{Triangular diagram with $\mathcal Q_{LL}^{(\prime\prime)}$ insertion contributing to $B_{d,s}-\bar{B}_{d,s}$ mixing. Anomalous vertex couples two quarks, $W$ and a would-be Goldstone boson or two quarks and two would-be Goldstone bosons.}
\label{slika2}
\end{center}
\end{figure}

Finally, in the general $R_\xi$ gauge $\mathcal Q^{(\prime\prime)}_{LL}$ operators contribute to $B_{d,s}-\bar{B}_{d,s}$ mixing amplitude also through triangle diagrams presented in Fig.~\ref{slika2}. 

We only consider single insertions of all the operators. This is a good approximation given the small size of observed deviations in the CP-conserving $B_{d,s}$ mixing observables from SM predictions. However, we have also computed higher order insertions and checked explicitly that they do not change our conclusions of the numerical analysis presented in the next section.

By simple consideration of the chirality structure of the diagrams we find that {single insertions of operators $\mathcal Q_{RR}$ and $\mathcal Q_{LRb}$ give contributions suppressed by the external down quark masses. Moreover, contributions of the two operators to the $B\to X_s \gamma$ decay rate have been considered in Ref.~\cite{Grzadkowski:2008mf}. Although there the NP contributions were assumed to be real, the resulting constraints on $C_{RR}$ and $C_{LRb}$ are very severe due to a parametric $m_t/m_b$ enhancement of these contributions compared to the SM result. Consequently, the operators $\op_{RR}$ and $\op_{LRb}$ are precluded from contributing significantly to $B_{d,s}$ mixing observables and will not be considered further.  }

After neglecting the masses of $u$ and $c$ quarks and enforcing unitarity of the CKM matrix we obtain the LO Wilson coefficients of effective Lagrangian in (\ref{eq:lagrBB})
{
\begin{eqnarray}
\nn\Delta C_1 &=&\{\mathrm{Re}[\kappa_{LL}] +  \kappa_{LL}^{\prime\prime}/2 \} S_0^{LL}(x_t,\mu) + 2  \kappa_{LL}^{\prime}  S_0^{\rm SM}(x_t)\nonumber\\
&& +\{\mathrm{Re}[\kappa_{LRt}] +  \kappa_{LRt}^{\prime}/2 \}S_0^{LRt}(x_t) + \kappa_{\rm c.t.}(\mu)\,,
\label{LOWils}
\end{eqnarray}
}
where $x_t=m_t^2/m_W^2$, $C_{1} = C_1^{\mathrm{SM}}+ \Delta C_1$ (at LO $C_1^{\rm SM} = S_0^{\rm SM}(x_t)$) and $\kappa_i$ are
\begin{align}
\kappa_{LL}^{(\prime,\prime\prime)}&=\frac{C_{LL}^{(\prime,\prime\prime)}}{\Lambda^2\sqrt{2}G_F}\,,& \kappa_{LRt}^{(\prime)}&=\frac{C_{LRt}^{(\prime)}}{\Lambda^2 G_F}\,.
\label{eq:kappas}
\end{align}
The $S_0^i(x_t)$ functions can be found in the Appendix~\ref{sec:App2}. Their gauge independence has been checked by the cancelation of all $\xi$-dependent terms. {On the other hand, $S_0^{LL}$ contribution turns out to be UV-divergent. We renormalize it using the $\overline{\mathrm{MS}}$ prescription, leading to a remnant  $\log \mu^2/m_W^2$ renormalization scale dependence. 
Because of this ultraviolet renormalization, it would be inconsistent to assume that no other operators but those in (\ref{eq:ops1}) and (\ref{eq:ops2}) are present in the dimension-six part of the Lagrangian (\ref{eq:lagr}). In particular, on dimensional grounds it is easy to verify that the appropriate MFV consistent counter-terms are represented by the four-quark operators of the form $ [\bar Q \mathcal A_{QQ}\gamma_\mu Q]  [\bar Q \mathcal A'_{QQ} \gamma^\mu Q] $, where $\mathcal A_{QQ}^{(\prime)}$ are polynomials of $Y_u{Y_u}^\dagger$ and/or $Y_d {Y_d}^\dagger$. Generic tree-level contributions of this kind to $\Delta C_1$ in (\ref{LOWils}) are denoted by $\kappa_{\rm c.t.}$. They have been analyzed in detail in~\cite{Ligeti:2010ia} although not in the context of  radiative corrections but as standalone dimension-six $\Delta F=2$ effective operators adhering to MFV -- we will not consider them further\footnote{A discussion of the most general set of dimension-six operators, which can serve as counter-terms for radiatively induced $\Delta F=2$ transitions can be found in~\cite{BM}.}.  It is however important to keep in mind that 
our derived bounds on $\kappa_i$ presented in the next section assume that the dominant NP effects at the $\mu\simeq m_t$ scale are represented by a single $\kappa_i$ insertion.
}



Note that only real parts of $\kappa_{LL}$ and $\kappa_{LRt}$ enter Eq.~(\ref{LOWils}) and thus cannot introduce a new CP violating phase. On a computational level, this is due to the fact that these operators always contribute to the mixing amplitudes in hermitian conjugate pairs (affecting $tWb$ and $tWs/d$ respectively) always preserving the CKM flavor and CP structure.  It can also be understood more generally already at the operator level. Namely as shown in \cite{Blum:2009sk}, a necessary condition for new flavor violating structures $\mathcal Y_x$ to introduce new sources of CP violation in quark transitions  is that ${\rm Tr}(\mathcal Y_x[\langle Y_u Y_u^\dagger \rangle , \langle Y_d Y_d^\dagger \rangle])\neq 0$. In MFV models (where $\mathcal Y_x$ is built out of $Y_u$ and $Y_d$ ) this condition can only be met if $\mathcal Y_x$ contains products of both $Y_u$ and $Y_d$. In our analysis this is true for all operators except $\mathcal Q_{LL}$ and $\mathcal Q_{LRt}$.

\section{Numerical results}
In order to evaluate the hadronic matrix elements of the operators, we evolve the Wilson coefficients (\ref{LOWils}) from the matching scale at the top quark $\overline{\rm MS}$ mass $m_t\equiv\overline{m}_t(\overline{m}_t)$ to the low energy scale at bottom quark $\overline{\rm MS}$ mass $m_b\equiv\overline{m}_b(\overline{m}_b)$. Using results given in Ref.~\cite{Buras:2001ra} we perform the next-to-leading log (NLL) running in the $\overline{\mathrm{MS}}$(NDR) scheme and obtain
\begin{eqnarray}
C_1(m_b)&=& 0.840\, C_1(m_t)\,.
\end{eqnarray}
We note that, since our weak scale matching is only done at LO in QCD, there is an ambiguity of the order $\alpha_s(m_t)/4\pi$ and a residual scheme dependence, when performing the RGE evolution at NLL. 
However $\mathcal (\alpha_s)$ corrections to the matching are in general model dependent and thus beyond the scope of our effective theory approach (c.f.~\cite{Becirevic:2001jj} for a more extensive discussion on this point).

In order to be consistent with the normalization of the SM contributions in Ref.~\cite{Lenz:2010gu}, from where we take the allowed ranges for $\Delta_{d,s}$, we use numerical values of $m_{b,t}$ and the hadronic matrix elements of $\op^{d,s}_1$ specified therein and obtain
\begin{eqnarray}
\nn\Delta_{d,s}&=&1- 2.57\, \mathrm{Re}[\kappa_{LL}]-1.54\,\mathrm{Re}[\kappa_{LRt}]\\
&+&2.00\, \kappa_{LL}^{\prime}-1.29\, \kappa_{LL}^{\prime\prime} - 0.77\,\kappa_{LRt}^{\prime}\,.\label{num}
\end{eqnarray}

Using Eq.~(\ref{num}), we consider one $\kappa_i(\mu=m_t)$ at the time to be non-zero. The absence of a new CP violating phase in $\mathcal Q_{LL}$ and $\mathcal Q_{LRt}$ contributions makes them fall under the ``scenario II'' of \cite{Lenz:2010gu} resulting in the following bounds
\begin{align}
-0.082 < &\mathrm{Re}[\kappa_{LL}]<0.078 \,,&\text{at 95\% C.L.}\,,\\
-0.14 < &\mathrm{Re}[\kappa_{LRt}]<0.13 \,,&\text{at 95\% C.L.}\,.
\end{align}
Compared to existing $B\to X_s \gamma$ constraints given in Ref.~\cite{Grzadkowski:2008mf}, we find our bounds on $\kappa_{LL}$ to be comparable, while bounds on $\kappa_{LRt}$ are considerably improved. 

Contributions of other operators can contain new CP violating phases
and thus fall under the ``scenario III'' of \cite{Lenz:2010gu}. We present the resulting best-fit values of the corresponding $\kappa$'s in Table~\ref{tabFIT}.  
\begin{table}
\begin{center}
\begin{tabular}{c||r|r||}
\multicolumn{1}{c||}{}&\multicolumn{1}{c}{Re} & \multicolumn{1}{|c||}{Im}\\\hline
$\kappa_{LL}^{\prime}$&$-0.062^{+0.063}_{-0.030}$ & $-0.110^{+0.029}_{-0.024}$\\
$\kappa_{LL}^{\prime\prime}$&$0.097^{+0.048}_{-0.098}$ &$0.180_{-0.044}^{+0.037}$\\
$\kappa_{LRt}^{\prime}$&$0.160^{+0.079}_{-0.160}$ & $0.290^{+0.062}_{-0.074}$\\
\end{tabular}
\end{center}
\caption{Best-fit values for real and imaginary parts of $\kappa_i$ parameters and 1$\sigma$ C.L. intervals. }
\label{tabFIT}
\end{table}
These three operators were not considered in~\cite{Grzadkowski:2008mf}, however using their formulae it is straight forward to check that at least for purely real contributions, $\kappa^{\prime(\prime\prime)}_{LL}$ are not overly constrained by the $B\to X_s\gamma$ decay rate measurement\footnote{In particular, $\op'_{LL}$ contributions are exactly proportional to the LO SM calculation, while $\op''_{LL}$ effects are just one half of those by $\op_{LL}$. }. 
A more conclusive comparison of all the different indirect bounds on these effective operators and especially $\op'_{LRt}$ is beyond the scope of this paper but is in progress.

\begin{figure}[h]
\begin{center}
\includegraphics[scale= 0.6]{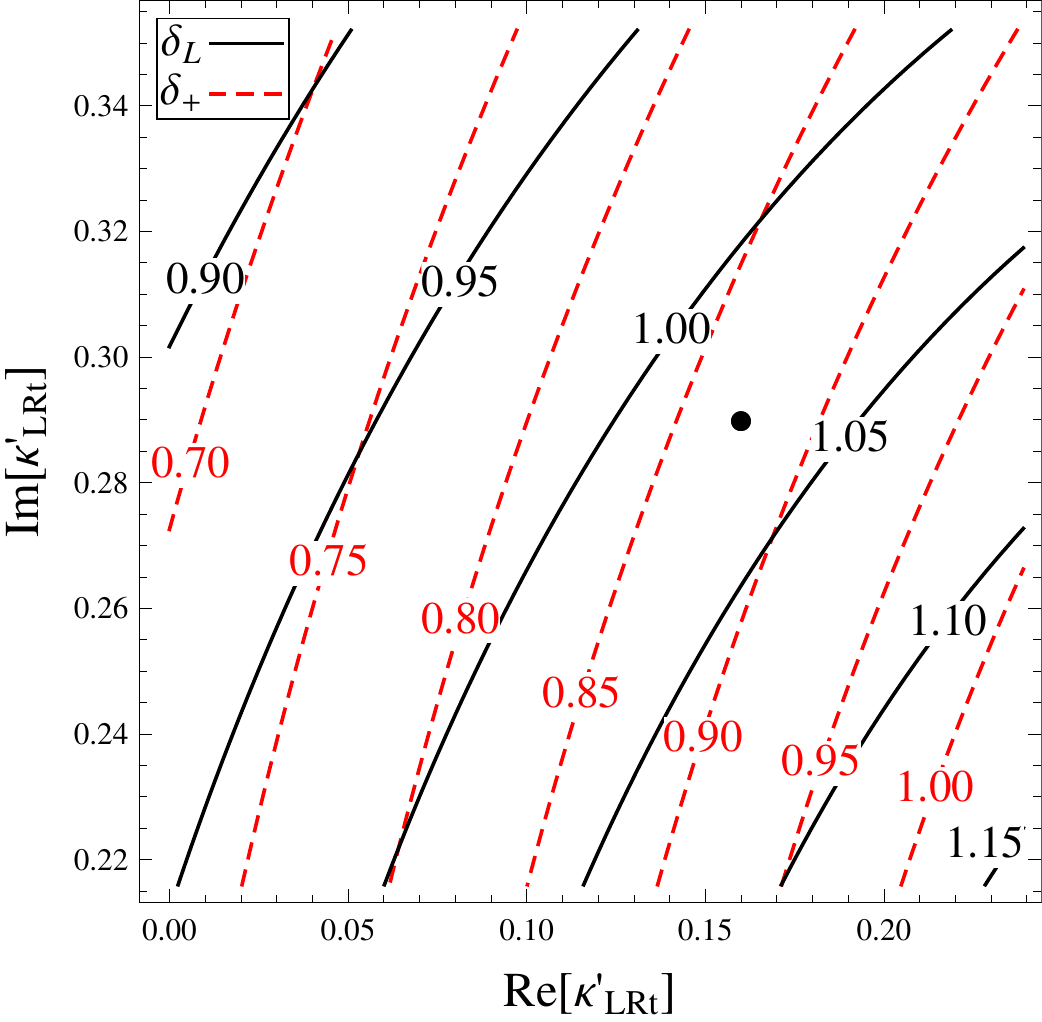}
\caption{Contour plot of $\delta_{L}\equiv{\cal F}_L/{\cal F}_{L}^{\mathrm{SM}}$ (black, solid) and $\delta_+\equiv{\cal F}_+/{\cal F}_+^{\mathrm{SM}}$ (red, dashed) as a function of real and imaginary part of $\kappa_{LRt}^{\prime}$ varied within the 1$\sigma$ C.L. interval given in Table~\ref{tabFIT}. Dot marks the position of the best-fit value for $\kappa_{LRt}^{\prime}$. {Fixing $C'_{LRt} = \exp(i\sigma)$, it corresponds to the NP scale and CP violating phase values of $\Lambda=0.51$\,TeV and $\sigma=61^\circ$.}}
\label{interplay3}
\end{center}
\end{figure}

Finally, with these results at hand, we reconsider the effects of our effective operators in Eq.~(\ref{eq:lagr}) on the helicity fractions of the $W$ boson in the main decay channel of the top quark, provided these same operators are responsible for new CP violating contributions in $B_{d,s}$ meson mixing. Both $\op_{LL}^{\prime(\prime\prime)}$ have the same chiral structure as the SM contribution and thus cannot affect the helicity fractions. They only yield small corrections to the total $t\to b W$ decay rate. On he other hand $\op'_{LRt}$ contributes in the same way as $\op_{LRt}$ and using the results obtained in Ref.~\cite{Drobnak:2010ej} we compute its effect on the $W$ boson helicity fractions (${\cal F}_{L,+}$) in the $t\to b W$ decay when the corresponding $\kappa'_{LRt}$ is varied within the $1\sigma$ C.L. region in Table~\ref{tabFIT}. The results are shown in Fig.~\ref{interplay3}. 
Compared to the SM predictions, ${\cal F}_{L,+}$ can deviate by as much as $15\%$ and $30\%$ respectively, although much smaller deviations are perfectly consistent with the ranges of $\kappa'_{LRt}$, preferred by $B_{d,s}$ mixing analysis. A robust prediction that can be made however is that at least one of the two independent helicity fractions (${\cal F}_{L,+}$) needs to deviate by at least $5\%$ from the corresponding SM prediction. While this is clearly beyond the reach of the LHC experiments for the ${\mathcal F}_+$, it is comparable to the expected precision for ${\mathcal F}_L$~\cite{AguilarSaavedra:2007rs}.

\section{Conclusions}

Within the framework of a weak scale MFV effective theory we have constructed a set of dimension $\leq 6$ effective operators  describing anomalous $ t W d_j$ interactions.  In the limit where the effects of multiple bottom yukawa insertions are neglected, we recover the set of operators previously considered in the study of the $B \to X_s \gamma$ decay. Taking into account possible large bottom Yukawa effects introduces additional operators with distinct new flavor structures. In particular, anomalous $ t W b$ interactions with either $t$ and/or $b$ left-handed can effectively be decoupled from those involving the first two quark generations. 

We have found that seven of the considered operators can possibly give sizable contributions to the $B_{d,s} - \bar B_{d,s}$ mixing amplitudes. Of those, five can also provide new sources of CP violation. Following the recent analysis of the CKMFitter group we have derived preferred ranges for the corresponding Wilson coefficients.  Several of the derived constraints improve upon previous bounds coming from the $B \to X_s \gamma$ analysis or are consistent with them. 



Finally, 
we find that in the presence of such new $ t W d_j$ interactions, the $W$ helicity fractions  ${\cal F}_{+,L}$ in the $t\to b W$ decay can deviate by as much $30\%, 15\%$ with respect to their SM values. The latter modification is within the expected precision of the LHC experiments.

%
%
%
%
%

\begin{acknowledgments}
J.F.K acknowledges enlightening discussions with Gilad Perez, Alex Kagan and Gino Isidori, and thanks the Weizmann Institute of Science and INFN Laboratori Nazionali di Frascati where part of this work was completed for their hospitality.
This work is supported in part by the European Commission RTN  network, Contract No. MRTN-CT-2006-035482 (FLAVIAnet) and by the Slovenian Research Agency. 
\end{acknowledgments}

\appendix

\section{Feynman rules\label{sec:App1}}
In Table~\ref{Tab2} we list the Feynman rules relevant for our analysis. 
\begin{table}[h!]
\begin{tabular}{m{1.8cm}|ll}
\includegraphics[scale= 0.5]{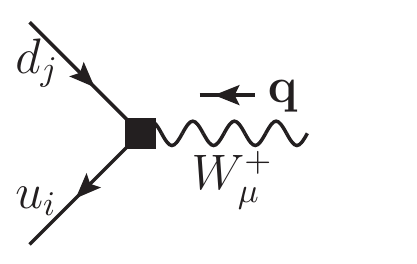}
 &$\frac{\ii g}{\sqrt{2}}V_{ij}\Big[\gamma^{\mu}v_{R,L} P_{R,L}+\frac{\ii \sigma^{\mu\nu}q_{\nu}}{m_W}g_{R,L} P_{R,L}\Big]$\\
\includegraphics[scale= 0.5]{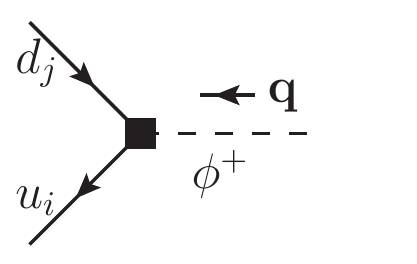}
 &$\frac{\ii g}{\sqrt{2}}V_{ij}\frac{\gs{q}}{m_W}v_{R,L} P_{R,L} $\\
\includegraphics[scale= 0.5]{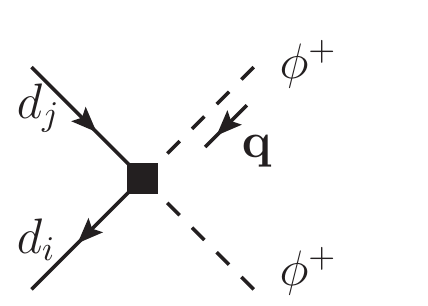}
 &$\Big(\frac{\ii g}{\sqrt{2}}\Big)^2(\kappa_{LL}+\kappa_{LL}^{\prime\prime})V_{ti}^* V_{tj}\frac{\ii \gs{q}}{m_W^2}P_L$\\
\includegraphics[scale= 0.5]{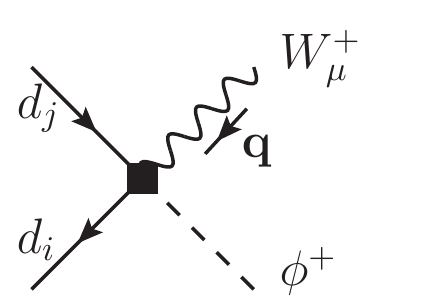}
 &$\Big(\frac{\ii g}{\sqrt{2}}\Big)^2(\kappa_{LL}+\kappa_{LL}^{\prime\prime})V_{ti}^* V_{tj}\frac{\ii}{m_W}\gamma^{\mu}P_L$\\
\end{tabular}
\caption{\small Feynman rules for relevant vertices, where $P_{L,R}=(1\mp\gamma^5)/2$. Indicies $i$ and $j$ label quark flavor and $v_R = \kappa_{RR}\delta_{it}\delta_{jb}$, $v_L= \kappa_{LL}\delta_{it}+\kappa_{LL}^{\prime}+\kappa_{LL}^{\prime\prime}\delta_{it}\delta_{jb}$, $g_R =\kappa_{LRb}\delta_{jb}$, $g_L=\kappa^*_{LRt}\delta_{it}+\kappa_{LRt}^{\prime*}\delta_{it}\delta_{jb}$. In addition to $\kappa_i$ given in Eq.~(\ref{eq:kappas}), we define $\kappa_{RR}=C_{RR}/(\Lambda^22\sqrt{2}G_F)$ and $\kappa_{LRb}=C_{LRb}/(\Lambda^2G_F)$. }
\label{Tab2}
\end{table}
\section{Analytical formulae\label{sec:App2}}
Below we present the analytical formulae for the $S_0^i$ loop functions.
\begin{eqnarray}
S_0^{\mathrm{SM}}&=&\frac{x_t(x_t^2-11 x_t+4)}{4(x_t-1)^2}+\frac{3x_t^3\log x_t}{2(x_t-1)^3}\,,\\
S_{0\overline{\mathrm{MS}}}^{LL}&=&-\frac{x_t \left(x_t^2+10 x_t+1\right)}{2 \left(x_t-1\right)^2} + x_t \log \frac{m_W^2}{\mu^2}\\
\nn   &+&\frac{x_t \left(x_t^3-3 x_t^2+12 x_t-4\right) \log x_t}{\left(x_t-1\right)^3}\,,\\
S_0^{LRt}&=&3\sqrt{x_t}\bigg[-\frac{x_t(x_t+1)}{(x_t-1)^2}+\frac{2x_t^2\log x_t}{(x_t-1)^3}\bigg]\,.
\end{eqnarray}


\begin{thebibliography}{99}
\bibitem{tbWexp}
  T.~Aaltonen {\it et al.} [ The CDF Collaboration ],
  Phys.\ Rev.\ Lett.\  {\bf 105 } (2010)  042002.
  [arXiv:1003.0224 [hep-ex]]; 
  V.~M.~Abazov {\it et al.}  [D0 Collaboration],
  arXiv:1011.6549 [hep-ex].
  
\bibitem{AguilarSaavedra:2008zc}
  J.~A.~Aguilar-Saavedra,
  Nucl.\ Phys.\  B {\bf 812} (2009) 181
  [arXiv:0811.3842 [hep-ph]].
\bibitem{Drobnak:2010ej}
  J.~Drobnak, S.~Fajfer, J.~F.~Kamenik,
  Phys.\ Rev.\  {\bf D82 } (2010)  114008.
  [arXiv:1010.2402 [hep-ph]].
  
\bibitem{Grzadkowski:2008mf}
  B.~Grzadkowski, M.~Misiak,
  Phys.\ Rev.\  {\bf D78 } (2008)  077501.
  [arXiv:0802.1413 [hep-ph]].
\bibitem{Lenz:2010gu}
  A.~Lenz, U.~Nierste, J.~Charles {\it et al.},
  [arXiv:1008.1593 [hep-ph]].

\bibitem{Deschamps:2008de}
  O.~Deschamps,
  [arXiv:0810.3139 [hep-ph]].
  
  \bibitem{betas}
V.~M.~Abazov {\it et al.} [D0 Collaboration], 
Phys.\ Rev.\ Lett.\ {\bf 101} (2008) 241801
[arxiv:0802.2255 [hep-ex]];
T.~Aaltonen {\it et al.} [CDF Collaboration], 
CDF public note {\bf 9458};
 G.~Punzi, 
 PoS {\bf EPS-HEP2009} (2009) 022, 
 [arxiv:1001.4886 [hep-ex]]. 

\bibitem{Abazov:2010hj}
  V.~M.~Abazov {\it et al.} [ D0 Collaboration ],
  Phys.\ Rev.\ Lett.\  {\bf 105}, 081801 (2010).
  [arXiv:1007.0395 [hep-ex]].
    
\bibitem{Lee:2008xs}
  J.~P.~Lee, K.~Y.~Lee,
  [arXiv:0809.0751 [hep-ph]].
\bibitem{Lee:2010hv}
  J.~P.~Lee and K.~Y.~Lee,
  arXiv:1010.6132 [hep-ph].

\bibitem{MFV}
  R.~S.~Chivukula and H.~Georgi,
  Phys.\ Lett.\  B {\bf 188} (1987) 99;
  A.~J.~Buras, {\em et al.}
  Phys.\ Lett.\  B {\bf 500} (2001) 161
  [arXiv:hep-ph/0007085];
  G.~D'Ambrosio, G.~F.~Giudice, G.~Isidori and A.~Strumia,
  Nucl.\ Phys.\  B {\bf 645} (2002) 155
  [arXiv:hep-ph/0207036].

\bibitem{GMFV}
  A.~L.~Kagan, G.~Perez, T.~Volansky and J.~Zupan,
  Phys.\ Rev.\  D {\bf 80}, 076002 (2009)
  [arXiv:0903.1794 [hep-ph]].

\bibitem{Antonelli:2009ws}
  M.~Antonelli {\it et al.},
  Phys.\ Rept.\  {\bf 494}, 197 (2010)
  [arXiv:0907.5386 [hep-ph]].
  
\bibitem{Hurth:2008jc}
  T.~Hurth, G.~Isidori, J.~F.~Kamenik and F.~Mescia,
  Nucl.\ Phys.\  B {\bf 808}, 326 (2009)
  [arXiv:0807.5039 [hep-ph]].

\bibitem{:2005ema}
  [ ALEPH and DELPHI and L3 and OPAL and SLD and LEP Electroweak Working Group and SLD Electroweak Group and SLD Heavy Flavour Group Collaborations ],
  Phys.\ Rept.\  {\bf 427}, 257-454 (2006).
  [hep-ex/0509008].
  
  
\bibitem{Becirevic:2001jj}
  D.~Becirevic, M.~Ciuchini, E.~Franco {\it et al.},
  Nucl.\ Phys.\  {\bf B634 } (2002)  105-119.
  [hep-ph/0112303].
  
\bibitem{Ligeti:2010ia}
  Z.~Ligeti, M.~Papucci, G.~Perez, J.~Zupan,
  Phys.\ Rev.\ Lett.\  {\bf 105}, 131601 (2010).
  [arXiv:1006.0432 [hep-ph]].

\bibitem{BM}
W.~Buchmuller and D.~Wyler, 
Nucl.\ Phys.\ B {\bf  268}, 621 (1986).


\bibitem{Blum:2009sk}
  K.~Blum, Y.~Grossman, Y.~Nir {\it et al.},
  Phys.\ Rev.\ Lett.\  {\bf 102}, 211802 (2009).
  [arXiv:0903.2118 [hep-ph]].
  
  \bibitem{Buras:2001ra}
  A.~J.~Buras, S.~Jager, J.~Urban,
  Nucl.\ Phys.\  {\bf B605 } (2001)  600-624.
  [hep-ph/0102316].

\bibitem{Becirevic:2001xt}
  D.~Becirevic, V.~Gimenez, G.~Martinelli {\it et al.},
  JHEP {\bf 0204 } (2002)  025.
  [hep-lat/0110091].

\bibitem{AguilarSaavedra:2007rs}
  J.~A.~Aguilar-Saavedra, J.~Carvalho, N.~F.~Castro {\it et al.},
  Eur.\ Phys.\ J.\  {\bf C53}, 689-699 (2008).
  [arXiv:0705.3041 [hep-ph]].

\end{thebibliography}
\end{document}